\documentclass[aps,pra,twocolumn,superscriptaddress,groupedaddress,showpacs,floatfix]{revtex4}
\usepackage[dvips]{graphicx}
\usepackage{bm} 

\begin{document}
\title{Spin-dependent inelastic collisions in spin-2 Bose-Einstein condensates}
\author{Satoshi Tojo}
\affiliation{Department of Physics, Gakushuin University,
Tokyo 171-8588, Japan}
\author{Taro Hayashi}
\affiliation{Department of Physics, Gakushuin University,
Tokyo 171-8588, Japan}
\author{Tatsuyoshi Tanabe}
\affiliation{Department of Physics, Gakushuin University,
Tokyo 171-8588, Japan}
\author{Takuya Hirano}
\affiliation{Department of Physics, Gakushuin University,
Tokyo 171-8588, Japan}
\author{Yuki Kawaguchi}
\affiliation{Department of Physics, University of Tokyo,
Tokyo 113-0033, Japan}
\author{Hiroki Saito}
\affiliation{Department of Applied Physics and Chemistry, University
of Electro-Communications, Tokyo 182-8585, Japan}
\author{Masahito Ueda}
\affiliation{Department of Physics, University of Tokyo,
Tokyo 113-0033, Japan}
\affiliation{ERATO Macroscopic Quantum Project, JST, 
Tokyo 113-8656, Japan}
%
%
\date{\today}

\begin{abstract}

We studied spin-dependent two-body inelastic collisions in $F=2$ $^{87}$Rb
Bose-Einstein condensates both experimentally and theoretically.
The $^{87}$Rb condensates were confined in an optical trap
and selectively prepared in various spin states
in the $F=2$ manifold
at a magnetic field of 3.0 G. 
The measured atom loss rates depend on the 
spin states of colliding atoms.
We measured two fundamental loss coefficients for
two-body inelastic
collisions with total spins of 0 and 2.
The loss coefficients determine the loss rates of all the spin pairs. 
The experimental results for mixtures of all spin combinations are
in good agreement with numerical solutions of 
the Gross--Pitaevskii equations
that include the effect of a magnetic field gradient.

\end{abstract}

%
%
%
%
%
%

\maketitle

\section{Introduction}
The field of cold collisions
has attracted extensive interest and 
has grown explosively
since the early days of atom cooling and trapping
\cite{Weiner}.
The development of novel techniques for cooling and manipulating atoms have
led to a deeper understanding of physics of collisions:
e.g., evaporative cooling enabled 
Bose-Einstein condensates (BECs) to be realized
and opened the field of ultracold collisions, 
while the evaporative cooling process 
itself relies on the nature of collisions.
The mean-field theory of BEC, which successfully describes the properties
of BEC, depends on the elastic
$s$-wave scattering length to characterize the atomic interaction energy
\cite{Weiner,Klausen}.
Inelastic collisions also play important roles, e.g.,
in cooling and retaining an atomic or molecular cloud in a trap
\cite{Soeding,ARobert,Goerlitz,Kuwamoto,Schmaljohann,Widera,Tojo,Myatt}.

In an optical trap, the spin degrees of freedom of atoms are liberated
enabling a rich variety of spinor BECs physics to be studied.
Spin-2 BECs have attracted much interest in recent years 
and there have been several experimental studies of spin-2 systems,
including investigations of their magnetic phases
\cite{Kuwamoto,Schmaljohann,Widera,Tojo},
of multiply charged vortices \cite{Isoshima}, 
and the phase separation between spin-2 and spin-1 BECs \cite{Hall,Mertes}. 
It is, however, difficult to observe stationary states of these phenomena 
because 
the $^{85}$Rb $F=2$ BEC has a negative scattering length,
and in $^{7}$Li, $^{23}$Na, $^{39}$K, $^{87}$Rb 
the $F=2$ states are in the upper hyperfine manifold.
Ultracold gases in the upper hyperfine states
decay to the lower hyperfine state.
$^{87}$Rb atoms are highly suitable for investigating the properties of a spin-2 BEC
because they have much lower inelastic collision rates than other species:
e.g., 
$^{87}$Rb has a loss rate of the order of $10^{-14}$cm$^3/$s,
which is much lower than that of $^{23}$Na \cite{Goerlitz}.

In this paper, we systematically investigate 
ultracold two-body inelastic collisions between Zeeman states of $F=2$ atoms
both experimentally and theoretically.
The experiment was conducted by creating an $F=2$ spinor $^{87}$Rb BEC 
in an optical trap and populating desired Zeeman sublevels by
radio-frequency transitions.
We measured the atom loss rates for various initial populations.
Spin-exchange collisions were negligible at a bias magnetic field of 3.0 G. 
By analogy with the scattering length in elastic collisions,
two-body inelastic collisions are described by two parameters,
$b_0$ and $b_2$,
which correspond to channels with the total spins of 0 and 2, respectively.
We experimentally determine these two parameters from the loss rates of 
single-component BECs of $|F=2,m_F=1\rangle$ and $|F=2,m_F=0\rangle$. 
We also measured the atom loss rates for two-component BECs 
of all possible sets of magnetic sublevels 
and we compared the results with our theoretical model.
We calculated the time evolution of the number of atoms in two-component BECs
using the values of $b_0$ and $b_2$ obtained by two methods:
the single-mode approximation (SMA)
and the Gross--Pitaevskii (GP) equations 
that include the effect of a magnetic field gradient.
The results obtained with the latter method 
agreed well with the experimental results.

\section{Theory of two-body inelastic collisions}

We first consider a system of spin-2 $^{87}$Rb atoms with no atom losses.
The Hamiltonian for this system is given by
\begin{equation}
 \hat H = \sum_{m = -2}^2 \int d\bm{r} \hat\psi_m^\dagger(\bm{r})
\left[ -\frac{\hbar^2}{2M} \nabla^2 + V_m(\bm{r}) \right]
\hat\psi_m(\bm{r}) + \hat H_{\rm int},
\end{equation}
where $\hat\psi_m$ and $V_m$ are the field operator and an external
potential for the hyperfine state $|F = 2, m \rangle$, and
$M$ is the mass of the $^{87}$Rb atoms.
The interaction Hamiltonian $\hat H_{\rm int}$ has the form \cite{Ueda}
\begin{equation}
\hat H_{\rm int} = \sum_{{\cal F} = 0, 2, 4} g_{\cal F} \sum_{{\cal M} =
 -{\cal F}}^{\cal F} \int d\bm{r} \hat A_{\cal FM}^\dagger(\bm{r}) \hat
 A_{\cal FM}(\bm{r}),
\label{hamil}
\end{equation}
where
\begin{equation}
 g_{\cal F} = \frac{4\pi \hbar^2 a_{\cal F}}{M},
\end{equation}
with $a_{\cal F}$ being the $s$-wave scattering length with the colliding
channel of total spin ${\cal F}$ and $\hat A_{\cal FM}$ is the
annihilation operator of two atoms with total spin $|{\cal F, M} \rangle$ defined as
\begin{equation}
 \hat A_{\cal FM}(\bm{r}) = \sum_{m, m' = -2}^2 C_{m m'}^{\cal FM}
\hat\psi_{m}(\bm{r}) \hat\psi_{m'}(\bm{r})
\end{equation}
with 
$C_{m m'}^{\cal FM} \equiv \langle {\cal F, M} | 2, m; 2, m'\rangle$
being the Clebsch--Gordan coefficient.

We take into account only the two-body inelastic loss,
since the two-body inelastic collision
is the dominant atom loss process for the upper hyperfine state of
$^{87}$Rb atoms.
We neglect the three-body losses and density-independent losses, such as
background-gas scattering and photon scattering. 
The two-body inelastic collision changes the hyperfine spin of either or
both of the colliding atoms from $F = 2$ to $F = 1$.
Consequently, the colliding atoms acquire a kinetic energy
corresponding to the hyperfine splitting, allowing them to escape from
the trap.

We write the master equation for the time evolution of the system with
two-body inelastic loss as
\begin{equation}
 \frac{\partial \hat\rho}{\partial t} = \frac{1}{i\hbar} [\hat H,
	\hat\rho] + \left( \frac{\partial \hat\rho}{\partial t} \right)_{\rm
	loss},
\label{master}
\end{equation}
where $\hat\rho$ represents the density operator for the atoms in the $F
= 2$ manifold.
The second term on the right-hand side of Eq.~(\ref{master}) describes
the atom loss, which must be rotationally invariant.
If we assume that the atoms after the two-body inelastic collision
immediately escape from the system and do not interact with the other atoms,
the atomic-loss part of the master equation can be written as~\cite{Gardiner}
\begin{eqnarray}
 \left( \frac{\partial \hat\rho}{\partial t} \right)_{\rm loss}  = 
	\sum_{{\cal F} = 0, 2} \frac{b_{\mathcal{F}}}{4} \sum_{{\cal M} = -{\cal F}}^{\cal
	F} \int d\bm{r} \biggl[ 2 \hat A_{\cal FM}(\bm{r}) \hat\rho \hat
  A_{\cal FM}^\dagger(\bm{r})
\nonumber \\
- \hat A_{\cal FM}^\dagger(\bm{r}) \hat
	A_{\cal FM}(\bm{r}) \hat\rho - \hat\rho \hat A_{\cal
	FM}^\dagger(\bm{r}) \hat A_{\cal FM}(\bm{r}) \biggr],
\label{loss}
\end{eqnarray}
where the constant $b_{\cal F}$ characterizes the loss rate for
colliding channel with total spin ${\cal F}$.
We note that Eq.~(\ref{loss}) is rotationally invariant.
If the dipolar decay is negligible, the total spin ${\cal F}$ before and
after the collision must be the same, and therefore, inelastic decay
through the ${\cal F} = 4$ channel is prohibited in Eq.~(\ref{loss}):
this is experimentally confirmed in the following section.

\begin{figure}
\begin{center}
\includegraphics[width=6.0cm]{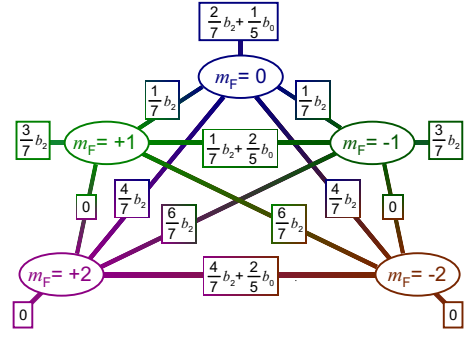}
\caption{(color online) Inelastic collision rates $K_2^{(m, m')}$
for all combinations of $m$ and $m'$ in a spin-2 BEC.
A rectangle that connects two ellipses corresponds to a collision
between two different components, 
whereas a rectangle linked to one ellipse
corresponds to a collision between the same component.
}
\label{map}
\end{center}
\end{figure}

For example, let us consider a mixture of $|2, 0\rangle$ and 
$|2, 1\rangle$ atoms.
Using Eq.~(\ref{loss}), we obtain the two-body inelastic loss of the
$|2, 0\rangle$ component as
\begin{eqnarray}
\left( \frac{\partial}{\partial t} \langle \hat\psi_0^\dagger
 \hat\psi_0 \rangle \right)_{\rm loss} & = & {\rm Tr}
\left[ \left( \frac{\partial \hat\rho}{\partial t} \right)_{\rm loss}
\hat\psi_0^\dagger \hat\psi_0 \right]
\nonumber \\
& = & -\left( \frac{1}{5} b_0 + \frac{2}{7} b_2 \right)
\langle \hat\psi_0^{\dagger 2} \hat\psi_0^2 \rangle
\nonumber \\
& & - \frac{1}{7} b_2 \langle \hat\psi_0^\dagger \hat\psi_0
\hat\psi_1^\dagger \hat\psi_1 \rangle,
\label{loss01}
\end{eqnarray}
where Tr represents the trace.
In the mean-field approximation, Eq.~(\ref{loss01}) is simplified as
\begin{equation}
 \frac{d n_0(\bm{r})}{dt} = -K_2^{(0, 0)} n_0^2(\bm{r}) - K_2^{(0, 1)}
	n_0(\bm{r}) n_1(\bm{r}),
\end{equation}
where $K_2^{(0, 0)} = b_0 / 5 + 2 b_2 / 7$, $K_2^{(0, 1)} = b_2 / 7$,
and $n_m$ is the density of component $m$.
In general, two-body inelastic decay in a two-component mixture has the form
\begin{equation}
 \frac{d n_{m}(\bm{r})}{dt} = -K_2^{(m, m)} n_{m}^2(\bm{r}) -
	K_2^{(m, m')} n_{m}(\bm{r}) n_{m'}(\bm{r}),
\label{loss02}
\end{equation}
and for a single-component system,
\begin{equation}
 \frac{d n_m(\bm{r})}{dt} = -K_2^{(m, m)} n_m^2(\bm{r}).
\label{eq2bodyloss}
\end{equation}
The coefficients $K_2^{(m, m')}$ for all $m$ and $m'$ are shown in
Fig.~\ref{map}.
%
%

From Eq.~(\ref{master}), the equation of motion for $\langle \hat\psi_m
\rangle$ is given by
\begin{equation}
 \frac{\partial}{\partial t} \langle \hat\psi_m(\bm{r}) \rangle = 
\frac{1}{i\hbar}
\langle [\hat\psi_m(\bm{r}), \hat H] \rangle
+ {\rm Tr} \left[\left(\frac{\partial \rho}{\partial t} 
\right)_{\rm loss} \hat\psi_m(\bm{r}) \right].
\label{psim}
\end{equation}
Using Eq.~(\ref{loss}), the second term on the right-hand side of
Eq.~(\ref{psim}) can be rewritten as
\begin{equation}
\sum_{{\cal F} = 0, 2} \frac{b_{\mathcal{F}}}{4} \sum_{{\cal M} = -{\cal F}}^{\cal
	F} \int d\bm{r}' 
\langle
[\hat A_{\cal FM}^\dagger(\bm{r}') \hat A_{\cal
	FM}(\bm{r}'), \hat\psi_m(\bm{r})] \rangle .
\label{aapsi}
\end{equation}
We find that Eq.~(\ref{aapsi}) has the same form as 
$\langle [\hat H_{\rm int},\hat\psi_m] \rangle$
with $g_{\cal F}$ being replaced by $b_{\cal F} / 2$.
In the mean-field theory, we thus expect that the two-body loss
can be incorporated into the 
GP equation by replacing
$g_{\cal F}$ with $g_{\cal F} - i \hbar b_{\cal F} / 2$.
Thus, the GP equation for a spin-2 BEC with two-body loss is
obtained as
\begin{eqnarray}
&& i \hbar \frac{\partial \psi_m}{\partial t} = \left( -\frac{\hbar^2
}{2M} \nabla^2 + V_m \right) \psi_m
\nonumber \\
\!\!\! + 
&&
\!\!\!\!\!\!\!\!\!
\sum_{{\cal F} = 0, 2, 4} 
\!\!\!
\tilde g_{\cal F} 
\!\!\!
\sum_{{\cal M} = -{\cal F}}^{\cal F} 
\sum_{m_1, m_2, m_3} 
\!\!\!\!\!\!
C_{m m_1}^{\cal FM} C_{m_2  m_3}^{\cal FM} 
\psi_{m_1}^* \psi_{m_2} \psi_{m_3},
\label{lossGP}
\end{eqnarray}
where
\begin{equation}
\tilde g_0 = g_0 - \frac{i \hbar b_0}{2}, \qquad
\tilde g_2 = g_2 - \frac{i \hbar b_2}{2}, \qquad
\tilde g_4 = g_4.
\end{equation}

If we assume that the atom density $n_m(\bm{r}, t)$ of a
single-component BEC always follows the Thomas-Fermi profile,
\begin{equation}
n_m(\bm{r}, t) = \frac{M}{4 \pi \hbar^2 a_m} [\mu(t) - V(\bm{r})],
\end{equation}
where $a_m$ is the scattering length for component $m$ and $\mu(t)$ is
determined from
\begin{equation}
N_m(t) = \int n_m(\bm{r}, t) d\bm{r},
\end{equation}
we can integrate Eq.~(\ref{eq2bodyloss}), giving~\cite{Soeding,Tojo}
\begin{equation}
\frac{dN_{m}(t)}{dt} = -\gamma_m K_{2}^{(m,m)} N_m^{7/5}(t).
\label{eq2bodyrho-1compo}
\end{equation}
Here the coefficient $\gamma_m$ is given by
\begin{equation}
\gamma_m = \frac{15^{2/5}}{14 \pi}
\left( \frac{M \bar{\omega}}{\hbar \sqrt{a_m}} \right)^{6/5}
\label{gamma}
\end{equation}
with $\bar{\omega}$ being the average trap frequency.
The solution of Eq.~(\ref{eq2bodyrho-1compo}) is obtained as
\begin{equation}
N_m(t) = \frac{5^{5/2}}{\left[5 N_m^{-2/5}(0) + 2 \gamma_m K_2^{(m,m)} t
\right]^{5/2}}.
\end{equation}
For a two-component system, we assume that both components $m$ and $m'$
have identical density profiles that are given by
\begin{eqnarray}
n_m(\bm{r}, t) & = & \frac{M}{4 \pi \hbar^2 \bar{a}} \rho_m(t)
 [\mu(t) - V(\bm{r})], \\
n_{m'}(\bm{r}, t) & = & \frac{M}{4 \pi \hbar^2 \bar{a}} \rho_{m'}(t)
 [\mu(t) - V(\bm{r})],
\end{eqnarray}
where $\bar{a}$ is the arithmetic average of the three scattering lengths
(between $m$-$m$, $m'$-$m'$ and $m$-$m'$), 
$\rho_{m (m')}(t) = N_{m (m')}(t) / N(t)$, 
and $\mu(t)$ is determined from
\begin{equation}
N(t) = \int \left[n_m(\bm{r}, t) + n_{m'}(\bm{r}, t) \right]d\bm{r}.
\end{equation}
We also assume that the coherent spin dynamics is negligible.
Such an SMA is accurate if the three
scattering lengths are the same.
Integrating Eq.~(\ref{loss02}), we obtain
\begin{equation}
\frac{dN_{m}(t)}{dt} =
-\bar{\gamma} N^{7/5}(t) \left[K_{2}^{(m,m)} \rho_{m}^2(t) 
+K_{2}^{(m,m')} \rho_{m}(t)\rho_{m'}(t) \right],
\label{eq2bodyrho}
\end{equation}
where $\bar{\gamma}$ has the same form as Eq.~(\ref{gamma}) in which $a_m$
is replaced by $\bar{a}$.

\section{Experimental setup}

The experimental apparatus and procedure to create $^{87}$Rb condensates are 
almost the same as in our previous study \cite{Kuwamoto, Tojo}.
A precooled thermal cloud prepared by the first magneto-optical trap (MOT) was
transferred to the second MOT by irradiating the first MOT
with a weak near-resonant
cw laser beam focused on the center of the MOT.
Over $10^9$ atoms in the second MOT were optically pumped into the
$|2,+2\rangle$ state and recaptured in an Ioffe-Pritchard (clover-leaf)
magnetic trap.
A BEC containing $10^6$ atoms was created by evaporative
cooling with frequency sweeping an rf field for 18 s.

The BEC was loaded into a crossed far-off-resonance optical trap (FORT) 
at a wavelength of 850 nm.
One of the laser beams propagated along the long axis of the
cigar-shaped BEC and had a power of 7 mW and 
a $1/e^2$ radius at the focus of 31 $\mu$m.
The other laser beam had a power of 11 mW and a beam waist of 97 $\mu$m
and was focused on and crossed the first beam.
The potential depth of the crossed FORT was estimated to be about 1 $\mu$K.
Both lasers had power fluctuations of less than 1\%.
The radial (axial) trap frequency measured from the parametric resonance
\cite{Friebel} was $2\pi\times 162$ Hz ($2\pi\times 21$ Hz).
The average trap frequency determined by 
measuring the parametric resonance was $2\pi\times 82$ Hz,
which  was consistent with 
that determined by measuring the release energy \cite{StamperKurn}.

The condensed atoms in the $|2,+2\rangle$ state in the magnetic trap
were transferred to the crossed FORT.
The magnetic trap was quickly turned off
and the quantization axis was then nonadiabatically inverted.
The condensed atoms were kept in the trap for 200 ms.
In the way,
condensates containing $3.2\times10^{5}$ atoms
in the $|2,-2\rangle$ state were prepared in the crossed FORT.
Some of the condensed atoms
in the $|2,-2\rangle$ state can be transferred to 
desired spin states $|2,m_F\rangle$
by the Landau-Zener transition
with an external magnetic field of 20.5 G \cite{Mewes}.
In this magnetic field, the condensed atoms 
can be selectively transferred
to the desired target states by quadratic Zeeman shifts.
The relative populations of $|2,m_F\rangle$ states
were controlled by the rf field strength and the sweep rate.
The observed fluctuation in the number of atoms
was estimated to be 10\%,
which was due to variations in the initial number of BEC atoms
and in the rf-transfer rate between spin states.
After state preparation,
the 20.5-G field was turned off and an accurately controlled external
magnetic field of 3.0 G was immediately applied.
The crossed FORT was turned off at the end of the time evolution period,
and absorption imaging was performed after a free expansion 
time of 15 ms
to measure the population distribution of each spin component
by the Stern--Gerlach method \cite{StamperKurn}.
A magnetic field fluctuation of $\sim$10 mG was measured by observing
the transfer rate for
magnetic dipole transitions
and the residual gradient magnetic field was estimated
to be $\sim$30 mG/cm \cite{Kuwamoto}.

\section{Results and discussion}

%
%
\begin{figure}
\begin{center}
\includegraphics[width=6.0cm]{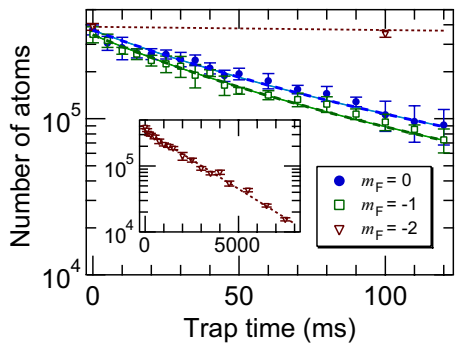}
\caption{(color online). Time dependence of 
the number of atoms in a single component BEC 
initially prepared
in $|2,0\rangle$ (filled circles), $|2,-1\rangle$ (open squares),
and $|2,-2\rangle$ (open triangles) states.
The inset shows the decay of $|2,-2\rangle$ atoms for a longer period. 
The experimental data are fitted by Eq.~(\ref{eq2bodyrho-1compo})
(dashed lines).
The theoretical curves are calculated from the GP equation (solid lines).
}
\label{atomloss}
\end{center}
\end{figure}

Figure \ref{atomloss}
plots the time dependence of 
the number of atoms of 
a single-component BEC
initially prepared in the $|2,0\rangle$ (filled circles),
$|2,-1\rangle$ (open squares),
and $|2,-2\rangle$ states (open triangles).
Each point represents an average typically over four samples. 
The observed condensate fraction was greater than 90\%
and the temperature of the thermal cloud was below 100 nK.
In the measurement, finite temperature effects in the condensates,
as seen in Ref.~\cite{SchmaljohannAPB}, were not observed.
The relative population of the initially prepared single component was
observed to be larger than 0.95 through the time evolution since
a magnetic field strength of 3.0 G suppressed the spin-exchange collisions
and only the initially prepared single component was observed through
the time evolution \cite{Kuwamoto}.
The quadratic Zeeman energies between $|2,0\rangle$ and $|2,\pm 1\rangle$
and between $|2,0\rangle$ and $|2,\pm 2\rangle$
were 31 nK and 124 nK, respectively.
The number of atoms in $|2,0\rangle$ and $|2,-1\rangle$ states
decreased rapidly with time.
By contrast, the number of atoms in the $|2, -2\rangle$ state remained almost
unchanged within 100 ms,
showing good agreement with the collision rate $K_2^{(-2,-2)}=0$.
The value of the one-body loss coefficient $K_1$ is experimentally estimated
to be $K_1 = 0.4$ s$^{-1}$ from 
the result for the $|2,-2\rangle$  state (see the inset of Fig.~\ref{atomloss}), 
and the three-body rate constant $K_3$ was deduced by 
S\"oding {\em et al.} \cite{Soeding}
to be $K_3 = 1.8\times 10^{-29}$ cm$^6$s$^{-1}$.
In our experimental condition, the contributions of 
the $K_1$ and $K_3$ terms 
to the atom loss are negligible compared with that of the $K_2$ term.
Therefore, the decays in the number of atoms for the $|2,0\rangle$ and
$|2,-1\rangle$ atoms are well fitted by Eq.(\ref{eq2bodyrho-1compo})
based on SMA calculation.
The inelastic collision rates are thus obtained as 
$K_{2}^{(0,0)} = (8.9\pm0.9)\times 10^{-14}$cm$^3$/s and 
$K_{2}^{(-1,-1)} = (10.4\pm1.0)\times 10^{-14}$cm$^3$/s
for $m_F = 0$ and $m_F = -1$, respectively.
From these values of $K_{2}^{(0,0)}=b_0 / 5 + 2 b_2 / 7$
and $K_{2}^{(-1,-1)}=3 b_2 / 7$, 
we deduce the inelastic collision coefficients 
$b_2 = (24.3\pm2.4)\times 10^{-14}$cm$^3$/s and 
$b_0 = (9.9\pm5.5)\times 10^{-14}$ cm$^3$/s
with scattering lengths of $94.57 a_B$ for $m_F=0$
and $95.68 a_B$ for $m_F=-1$
with $a_B$ being the Bohr radius,
respectively \cite{Widera}.
We also calculated the time evolution using the GP equations with the above values
to compare the results obtained using the GP equations with
those obtained using the SMA.
The difference between the results obtained by 
GP and SMA calculations is negligibly small as
shown in Fig.~\ref{atomloss}.

%
%
\begin{figure*}
\begin{center}
\includegraphics[width=11.0cm]{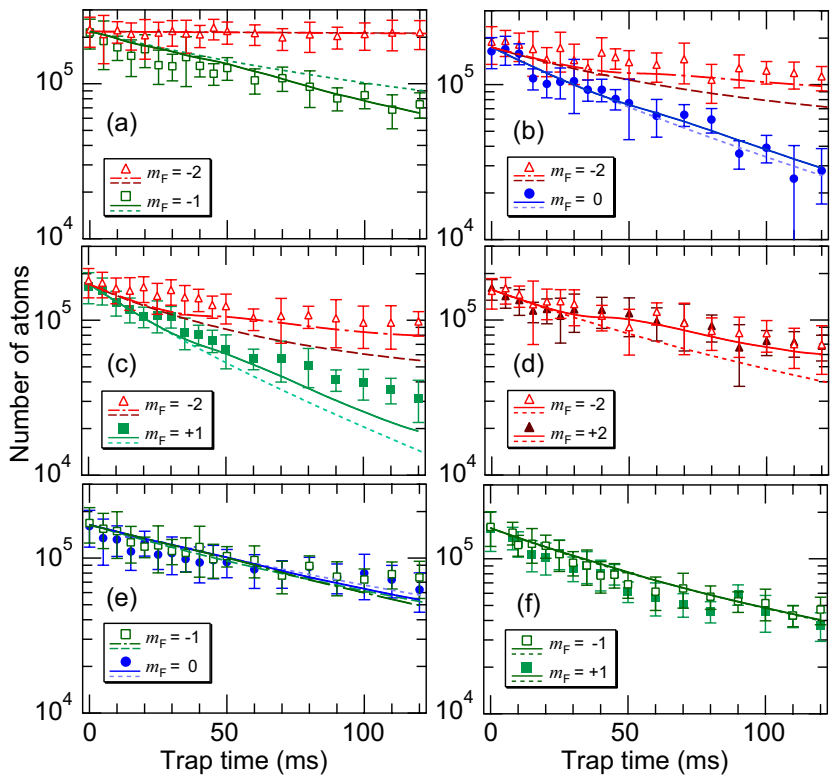}
\caption{(color online). Time dependence of numbers of condensed atoms for
various initially populated spin states
(a) $(2,-1) + (2,-2)$, 
(b) $(2,0) + (2,-2)$, 
(c) $(2,+1) + (2,-2)$,
(d) $(2,+2) + (2,-2)$,
(e) $(2,0) + (2,-1)$, and
(f) $(2,+1) + (2,-1)$ pairs.
The theoretical curves are calculated from Eq.~(\ref{eq2bodyrho}) with 
the SMA (dotted and dashed lines)
or the GP equations that include 
the effect of a magnetic field gradient (solid and dot-dashed lines).
}
\label{inespinall}
\end{center}
\end{figure*}
We next consider the loss of a two-component mixture to check the
consistency of Fig.~\ref{map}
combined with the measured values of $b_0$ and $b_2$.
We observed atom losses for all pairs of spin states
with initially equal populations as shown in Fig.~\ref{inespinall}.
The atom loss rates depend strongly on 
the spin states of colliding atoms.
We first discuss the atom loss rate 
for the $(2,-1) + (2,-2)$ pair.
As Fig.~\ref{inespinall}(a) shows,
the number of $|2,-2\rangle$ atoms remained almost unchanged
even in the presence of $|2,-1\rangle$ atoms.
This is consistent with the prediction $K_{2}^{(-1,-2)} =K_{2}^{(-2,-2)} =0 $ 
as shown in Fig.~\ref{map}.
On the other hand, 
for 
$(2,0) + (2,-2)$, 
$(2,+1) + (2,-2)$,
and 
$(2,+2) + (2,-2)$
pairs,
the $|2,-2\rangle$ component decreased with time 
as shown in Figs.~\ref{inespinall} (b), (c), and (d), respectively.
This decrease can be interpreted by two-body inelastic collisions 
of $|2,-2\rangle$
with $|2,0\rangle$, $|2,+1\rangle$, and $|2,+2\rangle$
 components because of the large values of 
$K_{2}^{(+1,-2)} = 6 b_2 / 7$,
$K_{2}^{(0,-2)} = 4 b_2 / 7$, and 
$K_{2}^{(+2,-2)} = 4 b_2 / 7 + 2 b_0 / 5$
from Fig.~\ref{map}.
The dotted and dashed lines are the SMA calculations
from Eq.~(\ref{eq2bodyrho})
using the measured values of $b_0$ and $b_2$,
an average trap frequency of $2\pi\times 82$ Hz, and
an average scattering length between spin states. 
The calculations are in qualitative agreement with the experimental results
for all pairs.
In particular, for pairs without the $|2,-2\rangle$ component 
(i.e. 
the $(2,0) + (2,-1)$  and
$(2,+1) + (2,-1)$  pairs 
shown in Figs.~\ref{inespinall} (e) and (f), respectively)
the SMA calculations with the Thomas--Fermi approximation
are in good agreement with the experimental results
indicating the prediction of Fig.\ref{map}.

However, with regard to the $|2,-2\rangle$ component,
the SMA calculations of 
the $(2,0) + (2,-2)$,
$(2,+1) + (2,-2)$,
and 
$(2,+2) + (2,-2)$ pairs are
slightly smaller than the experimental results as shown in
Figs.~\ref{inespinall} (b)--(d).
The decay curves for
$(2,+1) + (2,-1)$ and 
$(2,+2) + (2,-2)$ pairs are theoretically predicted to be identical
since 
$K_{2}^{(-1,-1)}+K_{2}^{(+1,-1)}=4 b_2/7+2 b_0/5$
is equal to $K_{2}^{(+2,-2)}$.
This is also understood from the fact that
$|2,+2\rangle + |2,-2\rangle$ is transformed to 
$|2,+1\rangle + |2,-1\rangle$ by spin rotation.
The SMA calculation results for $(2,+1) + (2,-1)$ pair 
are in good agreement with the experimental results.
However, the calculation for the $(2,+2) + (2,-2)$ pair
is smaller than the experimental results for trap times longer than 30 ms.

%
%
\begin{figure}
\begin{center}
\includegraphics[width=6.0cm]{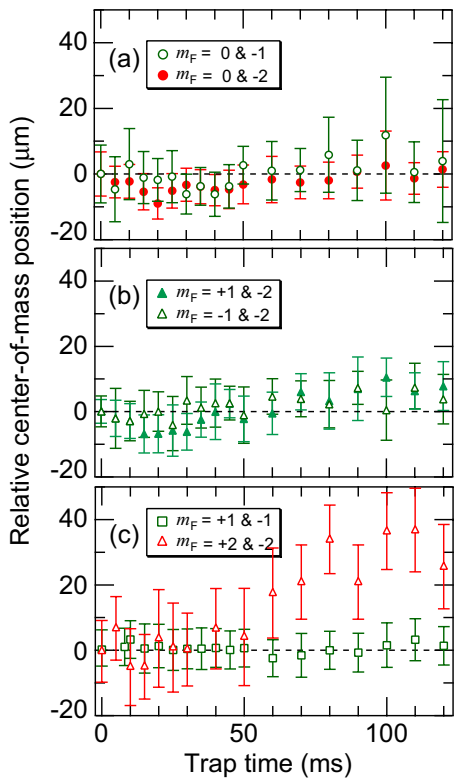}
\caption{(color online). Relative center-of-mass positions of
(a) $(2,0) + (2,-1)$ pair (open circles)
and $(2,0) + (2,-2)$ pair (closed circles),
(b) $(2,+1) + (2,-2)$ pair (closed triangles)
and $(2,-1) + (2,-2)$ pair (open triangles),
(c) $(2,+1) + (2,-1)$ pair (closed triangles)
and $(2,+2) + (2,-2)$ pair (open triangles).
}
\label{com4}
\end{center}
\end{figure}
Figure \ref{com4} shows relative center-of-mass position 
in axial direction ($z$-axis) of BECs between different spin states.
The center-of-mass position of $m_F=m$ atoms for a trap time $t$,
$z_m(t)$, was measured 
after Stern--Gerlach separation.
The separation between the center-of-mass position of $m_F=m$ and $m_F=m'$ atoms
due to the Stern-Gerlach magnetic field is $z_m(0) - z_{m'}(0)$ because
two components
overlapped at $t=0$.
The relative center-of-mass position for a trap time $t$ shown in
Fig. \ref{com4} is
given by $[z_m(t) - z_{m'}(t)] - [z_m(0) - z_{m'}(0)]$.
The magnetic field gradient could be the reason for
the difference in the relative center-of-mass positions between different spin states
since the low-field- and high-field-seeking states move in opposite directions.
The relative center-of-mass position for the $(2,+2) + (2,-2)$ pair
shifted to positive values with time while that for the $(2,+1) + (2,-1)$ pair 
leveled off almost zero.
The separation could reduce
the inelastic collision loss of the $(2,+2) + (2,-2)$ pair.

%
%
\begin{figure}
\begin{center}
\includegraphics[width=6.5cm]{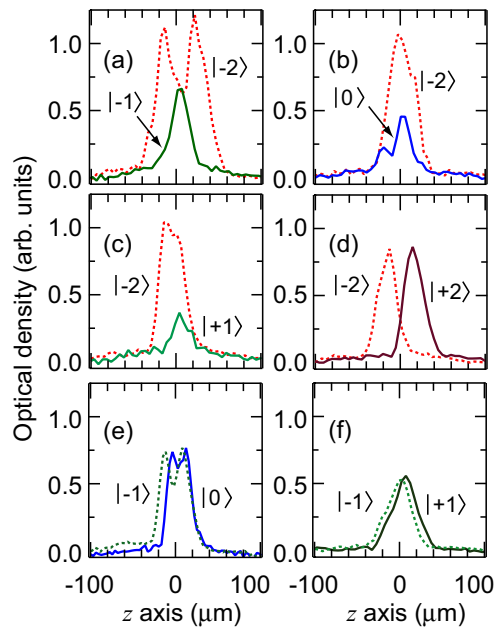}
\caption{(color online). 
Axial profiles of averaged density distributions for
various spin states at an evolution time of 100 ms.
Initially populated to
(a) $(2,-1) + (2,-2)$, 
(b) $(2,0) + (2,-2)$,
(c) $(2,+1) + (2,-2)$, 
(d) $(2,+2) + (2,-2)$, 
(e) $(2,0) + (2,-1)$, and
(f) $(2,+1) + (2,-1)$ states with equal population.
}
\label{odall}
\end{center}
\end{figure}

The separation may also be caused by mean-field interactions of BECs.
The miscibility of multicomponent BECs 
is determined by
$s$-wave scattering lengths
between identical and different spin states.
The interaction energy for a mixture of spin $m$ and $m'$ states is given by
\begin{equation}
E_{\rm{int}} = \int d \mbox{\boldmath$r$} 
   \left[ 
    \frac{g_{mm}}{2}  |\psi_m|^4 + \frac{g_{m'm'}}{2} |\psi_{m'}|^4 
    + g_{mm'} |\psi_m|^2 |\psi_{m'}|^2
   \right],
\label{eqIntE}  
\end{equation}
where
$g_{mm'}=4\pi \hbar a_{mm'}/M$ with $a_{mm'}$ being the scattering 
length between $m$ and $m'$ states. $a_{mm'}$ can be expressed in 
terms of $a_\mathcal{F}$ by comparing Eq.~(\ref{eqIntE}) and the mean-field 
expectation value of Eq.~(\ref{hamil}) for a mixture of $m$ and $m'$
states \cite{Saito}.
Two-component condensates is immiscible when
$g_{mm}g_{m'm'}<g_{mm'}^2$.
Therefore, based on the measured scattering lengths given in Ref. \cite{Widera}
the $(2,0) + (2,-1)$ and $(2,-1) + (2,-2)$ pairs 
are expected to have a phase separation 
while the other pairs are expected to be miscible.

Figure \ref{odall} shows
optical densities in the axial direction ($z$ axis)
of all spin-state pairs at a trap time of 100 ms,
which were averaged in the direction of gravity.
These density distributions depend on the spin states of the pairs.
Binary BECs of $(2,-1) + (2,-2)$ and
$(2,0) + (2,-1)$ pairs
exhibit phase separation and domain formation
as shown in Figs.~\ref{odall}(a) and (e),
due to the immiscibility condition.
The $(2,0) + (2,-2)$ and $(2,+1) + (2,-2)$ pairs
exhibit some inhomogeneity
between different spin states.
The $(2,+2) + (2,-2)$ pair exhibit phase separation 
despite it being predicted miscible on the basis of Eq.~(\ref{eqIntE})
and some other pairs have partial phase separation or homogeneous distribution in
Fig.~\ref{odall}.
This behavior could be explained by 
the effects of the magnetic field gradient.

The above analysis demonstrates that 
it is necessary to consider 
both the effect of the magnetic field gradient on motions in the FORT
and interactions between BECs in the theoretical analysis.
We numerically solved the GP equation (\ref{lossGP}) with 
$V_m = V_{\rm trap}+ B' \mu_B m z/ 2$, 
where the field gradient $B'$ in the axial direction
was assumed to be 30 mG/cm,
the radial (axial) trap frequency 
was $2\pi\times 162$ Hz ($2\pi\times 21$ Hz),
and the spin-dependent scattering lengths were derived from Ref.~\cite{Widera}.
The results are shown by the solid and dot-dashed lines in
Fig.~\ref{inespinall}.
These theoretical results are good quantitative agreement 
with the experimental results.
The differences between the theoretical and experimental results are 
significantly reduced for the
$(2,-1) + (2,-2)$, $(2,0) + (2,-2)$, and $(2,+2) + (2,-2)$ pairs
as shown Figs.~\ref{odall}(a), (b), and (d).
In particular, the calculation for the $(2,+2) + (2,-2)$ pair
can describe reduced atom losses after 50 ms
and a kink around 30 ms
as shown in Fig.~\ref{inespinall}(d).
However, 
the measured displacement of the relative center-of-mass position of
the $(2,+2)+(2,-2)$ pair
cannot be explained by magnetic field gradient alone.
We do not have an adequate theory for describing it quantitatively.
In the $(2,+1) + (2,-2)$ pair,
the theoretical result for $|2,+1 \rangle$ atoms 
cannot completely explain the experimental results,
while that for $|2,-2\rangle$ atoms is in better agreement with 
the experimental results.
This may be caused by the inhomogeneous density distribution.
The lower atom losses than the SMA calculation 
is due to the separation of the two components and the suppression of
two-body collision between them.

It has been theoretically predicted \cite{Ciobanu} and experimentally
measured \cite{Widera}
that the magnetic phase of a spin-2 $^{87}$Rb BEC is in close vicinity to
the boundary
between the cyclic and antiferromagnetic phases.
Saito and Ueda \cite{Saito} proposed a method to distinguish these two phases.
If the magnetic phase is cyclic, a $(2,+2)+(2,-2)$ mixture produces
$(2,0)+(2,0)$ pairs,
and observation of the $|2,0\rangle$ atoms in the $(2,+2)+(2,-2)$
mixture may be a signature of the cyclic phase.
However, the spin-dependent two-body losses are
not taken into account in Ref. \cite{Saito}.
To determine the magnetic phase for the upper hyperfine state,
it is necessary to take into account
the spin-dependent inelastic collision rates 
\cite{Tojo}.
Even if the $|2,0\rangle$ atoms are produced by the cyclic interaction, the
inelastic collision channels of
the $(2,0) + (2,+2)$ and $(2,0) + (2,-2)$ states 
may eliminate the $|2, 0\rangle$ atoms, thereby erasing the evidence of
the cyclic phase.
The obtained collision rates would enable us to quantitatively predict the
production rate of $|2,0\rangle$ atoms for the cyclic phase.
Another difficulty for determining the magnetic phase is 
the inhomogeneity of the system.
In Ref. \cite{Saito}, it is assumed that 
all condensates overlap homogeneously over the whole evolution time.
However,
in the measurement of the $(2,+2) + (2,-2)$ pair,
two components were separated in 
time evolution,
as shown in Fig.~\ref{odall}(d).
The magnetic phase in the ground state could be 
determined by spin population measurement
when condensate separation is prevented
by controlling the magnetic field gradient or
applying an additional external field such as optical lattices.

\section{Conclusions}

We have studied ultracold two-body inelastic collisions 
between Zeeman states of $F=2$ atoms
both experimentally and theoretically.
We measured spin-dependent atom loss rates for all possible 
combinations of spin states
by creating $F=2$ spinor $^{87}$Rb BECs in an optical trap and 
populating the desired Zeeman sublevels by radio-frequency transitions.
Atom loss rates were measured at a bias magnetic field of 3.0 G
at which spin-exchange collisions are negligible.
We show that the two-body inelastic collision rates between all Zeeman
sublevels can be 
expressed in terms of two inelastic collision constants, $b_0$ and $b_2$,
which respectively characterize collision channels with total spin 0 ($b_0$) 
and total spin 2 ($b_2$).
We have determined these values to be
$b_2 = (24.3\pm2.4)\times 10^{-14}$cm$^3$/s, and 
$b_0 = (9.9\pm5.5)\times 10^{-14}$ cm$^3$/s
from measured decay rates of single-component BECs
(i.e. the $|2,0\rangle$ and $|2,-1\rangle$ states).
Observed time evolutions of the atom number for two component BECs were consistent
with SMA calculations using the values of $b_0$ and $b_2$.
The discrepancy between the SMA calculations and the experimental results
can be explained by spatial separation between two components.
The calculated results from the GP equations with a magnetic field gradient
were in good agreement with the experimental results.

A detailed understanding of the relative-population dependence and 
the spin-state dependence of inelastic collisions are
key issues in the future study of spinor BECs, 
such as the determination of the magnetic ground state of spin-2 $^{87}$Rb BEC
and observation of novel quantum vortices.

\begin{acknowledgements}

We would like to thank T. Kuwamoto, E. Inoue, and Y. Taguchi for experimental
assistance and discussions.
This work was supported by 
Grants-in-Aid for Scientific Research
(Nos. 17071005, 19740248, and 20540388)
from the Ministry of Education, Culture, Sports, Science, and Technology
of Japan.
Additional funding was provided by CREST, JST.

\end{acknowledgements}


\begin{thebibliography}{99}
\bibitem{Weiner}
	As review, 
	J. Weiner, V.S. Bagnato, S. Zilio, and
	P.S. Julienne,
	Rev. Mod. Phys. \textbf{71}, 1 (1999).
\bibitem{Klausen}
	N.N. Klausen, J.L. Bohn, and C.H. Greene,
	Phys. Rev. A \textbf{64}, 053602 (2001).
\bibitem{Soeding}
	J. S\"oding,
	D. Gu\'ery-Odelin, P. Desbiolles, F. Chevy,
	H. Inamori, and J. Dalibard,
	Appl. Phys. B \textbf{69}, 257 (1999).
\bibitem{ARobert}
	A. Robert, 
	O. Sirjean, A. Browaeys, J. Poupard, S. Nowak,
	D. Boiron, C.I. Westbrook, and A. Aspect,
	Science \textbf{292}, 461 (2001)
\bibitem{Goerlitz}
	A. G\"orlitz, 
	T.L. Gustavson, A.E. Leanhardt, R. L\"ow,
	A.P. Chikkatur, S. Gupta, S. Inouye, D.E. Pritchard, and
	W. Ketterle,
	Phys. Rev. Lett. \textbf{90}, 090401 (2003).
\bibitem{Kuwamoto}
	T. Kuwamoto, K. Araki, T. Eno, and T. Hirano,
	Phys. Rev. A \textbf{69}, 063604 (2004).
\bibitem{Schmaljohann}
	H. Schmaljohann, 
	M. Erhard, J. Kronj\"ager, M. Kottke, S. van Staa, 
	L. Cacciapuoti, J.J. Arlt, K. Bongs, and K. Sengstock,
	Phys. Rev. Lett. \textbf{92}, 040402 (2004).
\bibitem{Widera}
	A. Widera, 
	F. Gerbier, S. F\"olling, T. Gericke, O. Mandel,
	and I. Bloch,
	New Journal of Physics \textbf{8}, 152 (2006).
\bibitem{Tojo}
	S. Tojo, A. Tomiyama, M. Iwata, T. Kuwamoto, and T. Hirano,
	Appl. Phys. B \textbf{93}, 403 (2008).
\bibitem{Myatt}
	C.J. Myatt,
	E.A. Burt, R.W. Ghrist, E.A. Cornell, and C.E. Wieman,
	Phys. Rev. Lett. \textbf{78}, 586 (1997),
	G. Modugno, 
	G. Ferrari, G. Roati, R.J. Brecha, A. Simoni, and
	M. Inguscio,
	Science \textbf{294}, 1320 (2001).
\bibitem{Isoshima}
	T. Isoshima,
	M. Okano, H. Yasuda, K. Kasa, J.A.M. Huhtam\"aki, M. Kumakura,
	and Y. Takahashi,
	Phys. Rev. Lett. \textbf{99}, 200403 (2007).
\bibitem{Hall}	
	D.S. Hall,
	M.R. Matthews, J.R. Ensher, C.E. Wieman, and E.A. Cornell,
	Phys. Rev. Lett. \textbf{81}, 1539 (1998).
\bibitem{Mertes}
	K.M. Mertes,
	J.W. Merrill, R. Carretero-Gonz\'alez, D.J. Frantzeskakis,
	P.G. Kevrekidis, and D.S. Hall,
	Phys. Rev. Lett. \textbf{99}, 190402 (2007).
\bibitem{Ueda}
	M. Ueda and M. Koashi,
	Phys. Rev. A \textbf{65}, 063602 (2002).
%
\bibitem{Gardiner}
	C. W. Gardiner and P. Zoller,
	\textit{Quantum Noise}, 2nd ed.
	(Springer, Berlin, 2000).
\bibitem{Friebel}
	S. Friebel,
	C. D'Andrea, J. Walz, M. Weitz, and T.W. H\"ansch
	Phys. Rev. A \textbf{57}, R20 (1998). 
\bibitem{StamperKurn}
	D. M. Stamper-Kurn, 
	M. R. Andrews, A. P. Chikkatur, S. Inouye,
	H.-J. Miesner, J. Stenger, and W. Ketterle,
	Phys. Rev. Lett. \textbf{80}, 2027 (1998).
\bibitem{Mewes}
	M.-O. Mewes, 
	M. R. Andrews, D. M. Kurn, D. S. Durfee,
	C. G. Townsend, and W. Ketterle,
	Phys. Rev. Lett. \textbf{78}, 582(1997).
%
\bibitem{SchmaljohannAPB}
	H. Schmaljohann, 
	M. Erhard, J. Kronj\"ager, 
	K. Sengstock, and K. Bongs,
	Appl. Phys. B \textbf{79}, 1001 (2004).
\bibitem{Ciobanu}
	C. V. Ciobanu, S.-K. Yip, and Tin-Lun Ho,
	Phys. Rev. \textbf{A} 61, 033607 (2000).
%
\bibitem{Saito}
	H. Saito and M. Ueda,
	Phys. Rev. A \textbf{72}, 053628 (2005).
%
\end{thebibliography}
%

\newpage

%
%
%

%
%
%

%
%
%


%
%
%

%
%
%

\end{document}